\documentclass[aps.prb,twocolumn,superscriptaddress]{revtex4-2}
\usepackage[a4paper,top=2.00cm,bottom=2.00cm,left=2.00cm,right=2.00cm]{geometry}
\usepackage{graphicx}
\usepackage{physics}
\usepackage{amsmath}
\usepackage{amssymb}
\usepackage{color}
\usepackage{hyperref}
\usepackage[version=4]{mhchem}
\usepackage{url}
\usepackage{siunitx}
\DeclareSIUnit\Ry{Ry}
\usepackage{multirow}
\begin{document}
\title{Trapping and Tunneling of Hydrogen, Deuterium and Oxygen in Niobium}
\date{\today}
\author{Abdulaziz Abogoda}
\affiliation{Hearne Institute for Theoretical Physics, Louisiana State University, Baton Rouge, LA 70803}
\affiliation{Applied Physics Graduate Program, Northwestern University, Evanston, IL 60208}
\author{J. A. Sauls}
\affiliation{Hearne Institute for Theoretical Physics, Louisiana State University, Baton Rouge, LA 70803}
\begin{abstract}
 
We investigate isolated O-H and O-D pairs trapped in BCC Nb using a machine-learning interatomic potential (MLIP) trained to density-functional theory (DFT). The MLIP enables large-supercell analysis and identification of trapping sites within BCC Nb, as well as efficient mapping of three-dimensional (3D) potential-energy surfaces. 
In addition to the pair of tetrahedral``face'' sites previously identified based on DFT, we identify a lower-energy pair of ``edge'' trapping sites and confirm the stability of H and D at these trapping sites with DFT. 
We solve the Schr\"odinger equation for H and D in the 3D potential that surrounds the trapping sites. Solutions based on the static-lattice limit yield tunnel splittings in the range $J/h \in\{3-100\}$ GHz for both trapping sites. 
\end{abstract}
\maketitle
\section{Introduction}

Niobium underpins a broad range of superconducting technologies: high-Q SRF cavities for high-energy accelerators~\cite{gra17}, and resonators and circuits for quantum computing and quantum sensing, searches for dark matter and physics beyond the Standard Model, and gravitational-wave detection~\cite{rom23,bog19,gao21,ber23,uek24}.

Microwave losses in niobium superconducting resonators operating at millikelvin temperatures are often attributed to ensembles of two-level tunneling systems (TLS)~\cite{mul19}.
Even after removal of surface oxides believed to host TLS defects, there is evidence of TLS defects embedded in the region of field penetration in Nb, i.e. the London penetration depth of order $\lambda_{L}\approx 50,\mathrm{nm}$~\cite{abo25,he25,rom20}.

Identifying the microscopic origin of TLSs in Nb devices, quantifying their impact on resonators and quantum circuits, and developing ways to suppress TLS-related loss and decoherence is central to advancing superconducting quantum processors and sensors~\cite{dev13,mul19}.
There is considerable experimental evidence from low-temperature heat capacity, acoustic and inelastic neutron studies implicating H and D atoms trapped near O or N interstitials as two-level tunneling systems~\cite{bir76,pfe76,mor78,wip81,mag83,wip84,bel85,mag86}.

In a recent study the authors utilized DFT and NEB methods to map the minimum energy path (MEP) for H tunneling between a pair of tetrahedral sites with O trapping at the octahedral site in the BCC unit cell of Nb. We analyzed the configuration proposed by Magerl et al.~\cite{mag83}, then solved the Schr\"odinger equation for the 1D MEP~\cite{abo25}. 
Our calculation of the tunnel splittings of the ground-state doublet for O-H and O-D  were found to be close to the tunnel splittings extracted from analysis of the low temperature heat capacity measurements by Wipf et al.~\cite{wip84} performed on Nb samples infused with very dilute concentrations of O–H and O–D ($x\simeq{0.05\%-1.4\%}$). These results are also consistent with inelastic neutron scattering~\cite{wip81}, quasi-elastic neutron scattering~\cite{ste91}, and ultrasonic attenuation~\cite{dre85}.

In this work we trained a high accuracy machine learning interatomic potential (MLIP) called MACE~\cite{bat22,bat22a}.
The MLIP enables large-supercell analysis and identification of trapping sites within BCC Nb, efficient mapping of potential-energy surfaces, as well as fast and accurate (compared to DFT) force calculations. 
In addition to the pair of tetrahedral ``face'' sites (the ``Magerl sites~\cite{mag83}) previously identified based on DFT, we find a lower-energy pair of ``edge'' trapping sites and confirm the stability of H and D at these trapping sites with DFT. 

We also utilized the MLIP to analyze the 3D trapping and tunneling of the O-H and O-D pairs with a static lattice and supercell dimensions using a method proposed in Ref.~\cite{sun04}.
In particular, we solve the Schr\"odinger equation for trapped H and D in the 3D potential that surrounds the trapping sites. Solutions based on the static potential yield tunnel splittings of $J_\mathrm{H}=0.414~\mathrm{meV}$ and $J_\mathrm{D}=0.024~\mathrm{meV}$ for the face sites, and $J_\mathrm{H}=0.275~\mathrm{meV}$ and $J_\mathrm{D}=0.014~\mathrm{meV}$ for the edge sites. The results for the face sites are close to our earlier results based on a nudged elastic band (NEB) determination of the 1D minimum energy path for H and D tunneling between face sites. The 3D results are also in good agreement with experimental reports of tunnel splittings for H and D trapped by O in Nb based on heat capacity and inelastic neutron scattering. A summary is provided in Table I.
The rest of the report discusses the MLIP and data base used to train the MLIP, the identification of trapping sites, calculation of the 3D potential energy surface (PES) and the analysis and calculations of the tunnel splittings for O-H and O-D for the two sets of trapping sites.

\section{MLIP training}

MACE is a higher-order equivariant message passing neural network potential \cite{bat22,bat22a}. We trained it to reproduce density-functional theory (DFT) energies, forces, and stresses. The training dataset comprises 4002 configurations. Every configuration is labeled with total energy and atomic forces; a substantial subset additionally includes the full stress tensor. To create a base for the model a collection of crystal structures involving combinations of Nb-O-H were evaluated, including relaxing distorted versions of the crystals. To achieve specificity for our use case the majority of the set spans variable and fixed cell relaxations of H and O interstitials in $3\times 3\times 3$ supercells of BCC Nb (individually and jointly), randomized starting placements, surfaces, and minimum-energy paths between interstitial sites obtained from nudged elastic band (NEB) calculations. Intermediate frames from relaxations and NEB images were also included. 

\begin{figure}
\includegraphics[]{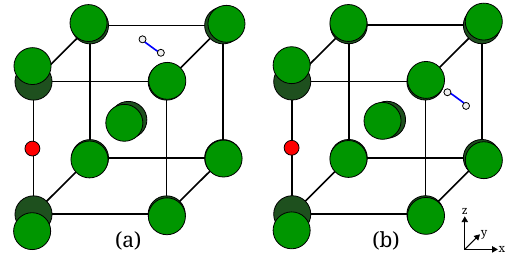}
\caption{(a) The Magerl or "face" site, and (b) the "edge" site. Red is O, gray is H, dark green is Nb in ideal BCC lattice, and light green are displaced Nb according to the method in section \ref{sec-splittings}.
\label{fig-configs}}
\end{figure}

The DFT software used to evaluate the configurations was QUANTUM ESPRESSO \cite{gia17,gia20}. Convergence parameters were chosen so that the energy per atom is converged to $\leq 1.36$ meV for every calculation. The electronic self-consistency tolerance was $1.36\times10^{-8}$ eV/atom. We used a plane-wave kinetic-energy cutoff of 1088 eV and Brillouin-zone sampling on Monkhorst–Pack meshes with target spacing of \SI{0.1}{\per\angstrom}, together with Marzari–Vanderbilt smearing of 0.054 eV. Exchange–correlation was treated within the Perdew-Burke-Ernzerhof (PBE) generalized-gradient approximation (GGA) \cite{per96}. For Nb and O we employed projector augmented-wave (PAW) pseudopotentials \cite{kuc14}; for H we used an optimized norm-conserving Vanderbilt (ONCV) pseudopotential \cite{sch15,ham13}. These settings were established by energy-per-atom convergence studies on Nb, NbO, and NbH unit cells and validated on Nb supercells with O-H interstitials.

All training and inference used float64 precision. Interatomic distances are expanded into a 10-dimensional set of smooth Bessel radial features up to  the cut off distance of \SI{6}{\angstrom}. We used correlation order of 3 (4 body interactions), 3 interaction (message-passing) layers, 128 channels, and maximum spherical order of 2. The loss function, combined energy, force, and (when available) stress terms; a multi-stage schedule gradually increased the relative weight on energies toward the end of training. An exponential moving average of the parameters with decay was maintained throughout optimization. The dataset was split 90/10 into training and validation sets. The model achieved mean absolute error in energy per atom $<$ 1 meV/atom in the validation set.

\section{Trapping Sites}

A big advantage of using fast and accurate MLIPs is the ability to evaluate and relax very large supercells. We made use of this to calculate the static trapping energy of H by O interstitials by constructing $8\times 8\times 8$ supercells. The baseline supercell was designed as an $8\times 8\times 8$ Nb supercell with one O interstitial and one H intersitial as far as possible from each other and fully relaxed to an energy with 1 meV convergence, we call this the ``far'' configuration. Then we systematically built and fully relaxed supercells of the same size with the O trapping H in all possible sites. The trapping energy for site $i$ was defined to be $E_{\text{trap},i} = V_i - V_\text{far}$. We highlight the 2 sites of interest. The first being the face configuration which the literature has so far focused on as the most probable trapping site with a trapping energy of \SI{-11}{\meV}. The ideal interstitial coordinates of this configuration is O at $a\,(0,\,0,\,0.5)$, with H tunneling between $a\,(0.75,\,0.5,\,1)$ and its mirror position across the $x=y$ plane. The second configuration is a newly identified site we call the edge sites. These have the lowest trapping energy at \SI{-62}{\meV}. For this config O is in the same position while H tunnels between $a\,(1,\,0.75,\,0.5)$ and its mirror position. Both are shown in Fig \ref{fig-configs}. We also confirmed the lower energy of the edge configuration by relaxing a $3\times 3\times 3$ supercell using DFT.

\begin{figure}
\includegraphics[]{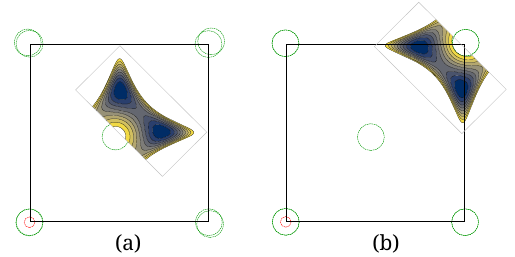}
\caption{Contour plots of the PES of H in (a) the face configuration and (b) the edge configuration. while centered on the bottom of the energy well in the z axis. Each contour represents \SI{50}{\meV}. The Gray box is the simulation box. The conventional cell is extracted from the center of the 4x4x4 supercell with the boundaries representing the dimensions of an unperturbed cell. Open circles mark the in-plane projection of atomic sites, green is Nb, red is O. 
\label{fig-Vs}}
\end{figure}

\section{Tunnel Splittings}\label{sec-splittings}

We compute energy levels and tunnel splittings, $J$, by solving the 3D time-independent Schrödinger equation on a sampled potential-energy surface (PES). 
The H nucleus is displaced over a rectilinear grid in the region of interest while total energies at each grid node are evaluated with our MACE model. For H and D the process is identical, as DFT and therefore the MLIP does not differentiate between isotopes. 
The difference in energy levels comes 
in the mass that enters the kinetic energy term in the Schrodinger equation.
These calculations are performed in $4\times 4\times 4$ supercell with the positions of the Nb lattice and the O impurity held constant as well as the cell vectors.

The static lattice is constructed in the manner similar to the method outlined in Ref.~\onlinecite{sun04}. We first find the relaxed position of H at a chosen site in the vicinity of O. We then perform a unique form of relaxation by which, in each relaxation step, the forces and energies which are used to find the minima are calculated by averaging the forces and energy of two supercells. The first with the H atom fixed at the position found in the first step and the second supercell with the H atom placed in its mirror position across the plane of symmetry (the x=y plane in both cases here). The end result gives a lattice that approximates the deformation around a superposition of H nuclei across the symmetric trapping sites. 

\begin{figure}
\includegraphics[]{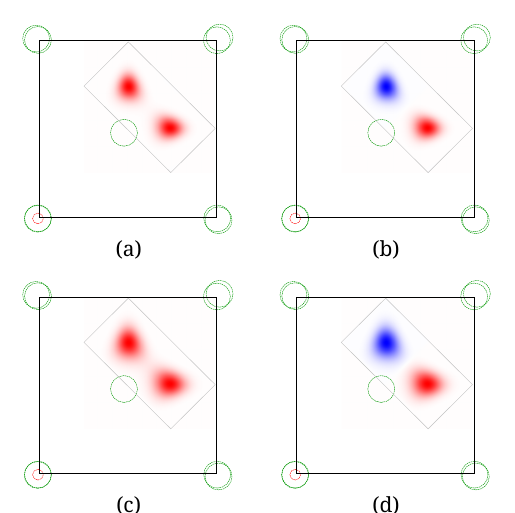}
\caption{Cross sections of wavefunctions (w.f.) of the face configuration $z=z_\text{max}$ were $z_\text{max}$ is the $z$ coordinate at which $|\phi|$ attains its maximum. Top (bottom) row is H (D). Left column shows the ground state (symmetric w.f.), while the right column shows the excited state (antisymmetric w.f.). Densities in panels: (a) $\phi_{\mathrm{H},0}$, (b) $\phi_{\mathrm{H},1}$, (c) $\phi_{\mathrm{D},0}$, and (d) $\phi_{\mathrm{D},1}$. The gray rectangle is the simulation box. The conventional cell is extracted from the center of the 4x4x4 supercell with the boundaries representing the dimensions of an unperturbed unit cell. Open circles mark the in-plane projection of atomic sites: green is Nb, red is O.
\label{fig-wf-T3}}
\end{figure}

The PES grid is then centered at the midpoint between the sites and is spans both energy wells with dimensions $(1.2, 1.2, 2.3)$ \unit{\angstrom} and rotated to capture the symmetry of the system. The grid resolution is parametrized by a vector N, specifying the number of grid points across each axis of the box. The PES can then be interpolated to finer or coarser resolution using piecewise cubic hermite interpolating polynomials (PCHIP) which was chosen for its continuity in value and first derivative and shape-preserving behavior and suppressing oscillations. Cut outs of the PES of both configurations are shown in Fig. \ref{fig-Vs}. Because our model is based on DFT data, we excluded grid points in which any pair of nuclei were separated by less than the sum of their respective pseudopotential cutoff radii, these grid points were assigned an arbitrary high potential value to simulate a hard wall.

\begin{figure}[t]
\includegraphics[]{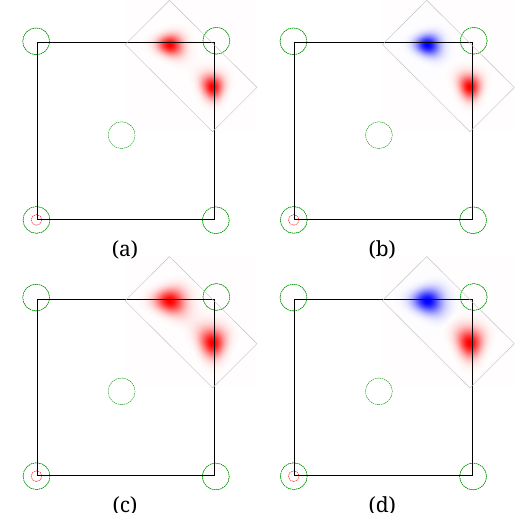}
\caption{Descriptions are the same as those in Fig. \ref{fig-wf-T3}, but for the edge configuration.
\label{fig-wf-T10}}
\end{figure}

We form a matrix Hamiltonian $H$ using a discretized version of the Schrodinger equation employing a central finite-difference for the Laplacian. After flattening the 3D grid to a 1D index using a standard 7-point stencil, the PES values populate the diagonal of $H$ and central finite difference factors generate the off-diagonal kinetic energy terms. Boundary grid points are excluded from $H$, which imposes a zero Dirichlet condition ($\phi(r)=0$) at the boundaries. The resulting eigenvalue problem is solved with a sparse Krylov-Schur solver with a required error of $<$ \SI{1}{\micro\eV}. The bare masses of H and D are used.

An initial PES is calculated using our MACE model and is solved successively for finer interpolated grids until in the error in energies or splittings vary by $<$ \SI{1}{\micro\eV}. As a final step a very fine grid is calculated using our MACE model in which convergence is achieved with less resolution than what the MACE PES provides, ensuring our convergence does not rely on interpolation. The final number of grid points for all calculations was $N=(51, 51, 102)$.

For the face configuration, we obtain tunneling splittings of
$J_\mathrm{H}=0.414~\mathrm{meV}$ ($100~\mathrm{GHz}$) and
$J_\mathrm{D}=0.024~\mathrm{meV}$ ($5.80~\mathrm{GHz}$).
For the edge configuration, the splittings are reduced to
$J_\mathrm{H}=0.275~\mathrm{meV}$ ($66.5~\mathrm{GHz}$) and
$J_\mathrm{D}=0.014~\mathrm{meV}$ ($3.39~\mathrm{GHz}$). 
Density plots of the the symmetric ground state $\phi_0$ and asymmetric first excited state wavefunctions $\phi_1$ for H and D are shown in Fig. \ref{fig-wf-T3} and Fig. \ref{fig-wf-T10} respectively. The isotope effect is observable when comparing spatial distributions of $\phi_\mathrm{H}$ and $\phi_\mathrm{D}$. 
Table \ref{table-summary} summarizes and compares results of 1D and 3D calculations as well as experimental measurments.

\begin{table}[]
\begin{tabular}{ll|c|c|ll}
\cline{3-4}
                                                           &                       & $J_\mathrm{H}$ [\unit{\meV}] & $J_\mathrm{D}$ [\unit{\meV}] &  &  \\ \cline{1-4}
\multicolumn{1}{|l|}{\textit{edge} sites}                  & 3D static lattice     & 0.275                        & 0.014                        &  &  \\ \cline{1-4}
\multicolumn{1}{|l|}{\multirow{2}{*}{\textit{face} sites}} & 3D static lattice     & 0.414                        & 0.024                        &  &  \\ \cline{2-4}
\multicolumn{1}{|l|}{}                                     & 1D VCNEB~\cite{abo25} & 0.358                        & 0.019                        &  &  \\ \cline{1-4}
\multicolumn{2}{|l|}{Experimental heat capacity}                                   & 0.190                        & 0.021                        &  &  \\ \cline{1-4}
\end{tabular}
\caption{DFT and MLIP based calculations of the H and D tunnel splittings for H and D atoms trapped by O in the Nb unit cell. The 1D calculations are based on DFT and NEB for the pair of face sites. The 3D calculations are based on the MLIP and the static trapping potentials for both face and edge sites. Results for H and D tunnel splittings from extremely low concentrations of O-H and O-D in Nb are shown for comparison.
\label{table-summary}
}
\end{table}

\section{Summary and Outlook}

We developed a high-accuracy Nb-O-H MACE model, trained using results based on DFT.  The model was first used to obtain the trapping energy of H and D by O in Nb. A new set of trappings sites were found, distinct from the face sites, with lower energy.
We then used the model to compute the energy levels and tunnel splittings of both configurations using the static lattice approximation. The static lattice increases the effective height of the tunnel barrier for the effective 1D MEP compared to that obtained using NEB in Ref.~\cite{abo25}, as can be seen in Fig.~\ref{fig-Vs}.
Furthermore, since the surrounding atoms are not adiabatically responding to the position of H or D there is no dynamically induced inertia and thus no effective mass correction.

In contrast to the static lattice calculation, NEB shows substantial adiabatic movement of the lattice in reaction to the position of H or D. Estimates of the timescales involved suggest that the static lattice approximation may not be sufficient. 
In particular, Nb phonon frequencies reach $\hbar\omega_D\sim\SI{28}{\meV}$~\cite{arn80,zar22}, which corresponds to a minimum timescale of $\sim$ \SI{100}{\fs}. The dimension of the supercell sets the minimum phonon frequency and therefore a timescale maximum, which for a $4\times 4\times 4$ supercell is $\sim\SI{600}{\fs}$. The timescale of TLS tunneling is set by $\sim \hbar/J$. The range of $J$ from \SI{0.2}{\meV} to \SI{1}{\meV} corresponds to timescales of $\sim$\SI{200}{\ps} to $\sim$\SI{4}{\ps}. Thus, the timescale for the tunneling process is orders of magnitude slower than the lattice relaxation process, suggesting an improved 3D calculation method would use an adiabatic potential energy surface (APES) and a method to account for the effective mass and coupling to phonons.

\section{Acknowledgments}

We thank William Shelton and Ilya Vekhter for discussions.
This work was supported by the Air Force Office of Scientific Research (AFOSR) under Award No. FA9550-23-1-0709
and 
the Hearne Institute of Theoretical Physics
at Louisiana State University.
Portions of this research were conducted with high performance computing resources provided by Louisiana State University (http://www.hpc.lsu.edu).


%
\end{document}